\pdfoutput=1
\documentclass[a4paper,aps,british,citeautoscript,pdftex,floatfix,groupedaddress,prb,twocolumn]{revtex4}

\usepackage{amsfonts,amsmath,amssymb}
\usepackage{amsmath, amssymb}
\usepackage{graphicx}
\usepackage[utf8x]{inputenc}
\usepackage{microtype}
\usepackage{pxfonts}
\usepackage{textcomp}
\usepackage{units}
\usepackage[hyperfootnotes=false]{hyperref}
\usepackage{hypernat}

\newcommand{\cm}{\ensuremath{\text{cm}^{-1}}}%
\newcommand{\eg}{e.~g.}%
\newcommand{\etal}{\emph{et~al}}%
\newcommand{\ie}{i.~e.}%
\newcommand{\JKaKc}{\ensuremath{J_{K_aK_c}}}%
\newcommand{\mueff}{\ensuremath{\mu_\text{eff}}}%
\newcommand{\ohstate}{\ensuremath{{}^2\Pi_{3/2},v=0,J=3/2}}%

\newlength{\figwidth}
\begin{document}
\preprint{Accepted for publication in \emph{Faraday Disc.} \textbf{142} (2009)}
\title{Manipulating the motion of large neutral molecules}%
\author{Jochen Küpper}%
\email{jochen@fhi-berlin.mpg.de}%
\author{Frank Filsinger}%
\author{Gerard Meijer}%
\affiliation{Fritz-Haber-Institut der Max-Planck-Gesellschaft, Faradayweg 4--6, 14195 Berlin,
   Germany}

\begin{abstract}
   \noindent%
   Large molecules have complex potential-energy surfaces with many local minima. They exhibit
   multiple stereo-isomers, even at very low temperatures. In this paper we discuss the different
   approaches for the manipulation of the motion of large and complex molecules, like amino acids or
   peptides, and the prospects of state- and conformer-selected, focused, and slow beams of such
   molecules for studying their molecular properties and for fundamental physics studies. \\
   Accepted for publication in \emph{Faraday Disc.} \textbf{142} (2009)
\end{abstract}
\renewcommand{\thefootnote}{\fnsymbol{footnote}}
\maketitle
\renewcommand{\thefootnote}{\arabic{footnote}}

\section{Introduction}
\label{sec:intro}

Over the last years there have been tremendous advances in the preparation of cold and ultracold
samples of small molecules, either by association of molecules from ultracold atoms, or by direct
cooling methods. Using the association technique, ultracold heteronuclear ground-state dimers were
recently produced~\cite{Ni:Science322:231, Deiglmayr:PRL101:133004}. Direct cooling methods allow
the preparation of trapped samples of more complex molecules (\ie,
ammonia~\cite{Bethlem:Nature406:491}) and they promise possibilities for their extension to large
molecules like the ``building blocks of life''~\cite{Weinkauf:EPJD20:309}. Recently, we have
demonstrated the alternating gradient (AG) deceleration of the prototypical large molecule
benzonitrile~\cite{Wohlfart:PRA77:031404}. Once such molecules are decelerated to a quasi-standstill
in the laboratory frame, they can be trapped in ac traps, which have been demonstrated for small
molecules in high-field-seeking states~\cite{Veldhoven:PRL94:083001, Schnell:JPCA111:7411}.

For many applications in physics and chemistry ensembles of large molecules all in one or in a few
quantum states would be highly beneficial. For many years, small molecules have been state-selected
and focused using static multipole fields~\cite{Ramsey:MolBeam:1956, Reuss:StateSelection}. For
about ten years it is also feasible to change the velocity of small molecules in low-field-seeking
states using the Stark decelerator~\cite{Bethlem:PRL83:1558}. The produced samples of slow molecules
can be trapped in static or dynamic fields, can be injected into storage rings, or can be used for
various molecular physics applications~\cite{Meerakker:NatPhys4:595}.

\begin{figure}
   \centering
   \includegraphics[width=\figwidth]{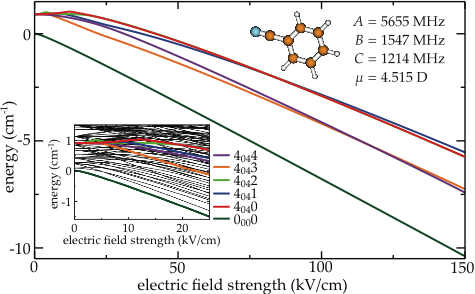}
   \caption{(Colour online): Energy of selected low rotational states of benzonitrile as a function
      of electric field strength. In the upper inset the molecular structure of benzonitrile and its
      relevant molecular parameters are given~\cite{Wohlfart:JMolSpec247:119}. In the lower inset,
      all states with a field-free energy below \unit[1.2]{\cm} are shown at smaller field
      strengths.}
   \label{fig:BN:Stark-curves}
\end{figure}

Larger molecules, however, have practically only high-field-seeking (hfs) states at the relevant
electric field strengths. To illustrate this the Stark curves of some low rotational states of
benzonitrile (C$_7$H$_5$N) are shown in figure~\ref{fig:BN:Stark-curves}.
In order to focus molecules in these hfs states one would need to create a maximum of electric field
in free space, which is not possible according to Maxwell's equations.\footnote{Nevertheless, the
   focusing of molecules in hfs states has been demonstrated using cylindrical capacitors with a
   central wire~\cite{Laine:PLA34:144} and arrays of crossed wires.\cite{Helmer:JAP31:458}} However,
large molecules have been deflected using static fields~\cite{Broyer:PhysScr76:C135,
   Holmegaard:PRL102:023001} and their motion was manipulated in that way in matter-wave
interferometry experiments~\cite{Berninger:PRA76:013607}. Moreover, the rotational motion of large
molecules has been restricted using brute-force orientation in dc electric
fields~\cite{Loesch:JCP93:4779, Friedrich:Nature353:412, Block:PRL68:1303}, using laser
alignment~\cite{Friedrich:PRL74:4623, Stapelfeldt:RMP75:543, Kumarappan:JCP125:194309}, or using
mixed dc and laser fields~\cite{Buck:IRPC25:583, Minemoto:JCP118:4052, Holmegaard:PRL102:023001}.

In order to confine the motion of large molecules, one has to use dynamic focusing in
alternating-gradient (AG) setups~\cite{Auerbach:JCP45:2160, Bethlem:JPB39:R263}. We have
demonstrated that alternating-gradient focusing can be used to focus and decelerate large molecules.
In a prototype experiment we have decelerated benzonitrile (C$_7$H$_5$N) molecules from a supersonic
jet~\cite{Wohlfart:PRA77:031404}. In similar experiments, we have demonstrated that the frequency
characteristics for the dynamic focusing in an AG setup can be used to separate species with
distinct mass-to-dipole moment ratios~\cite{Filsinger:PRL100:133003}. Equivalent to the
$m/q$-selection of charged particles in a mass spectrometer, these experiments perform an
$m/\mu$-selection. We have demonstrated the selection of the cis and trans conformers of
3-aminophenol (C$_6$H$_7$NO) and are currently performing first selection experiments on the
conformers of biomolecules. Slow, albeit warm, beams of thermally stable, large molecules can also
directly emerge from an oven~\cite{Deachapunya:EPJD46:307}.

The spectroscopic investigation of complex molecules isolated in the gas-phase has also seen big
advances over the last decade~\cite{Simons:PCCP6:E7, Vries:ARPC58:585}. These advances are largely
due to the ability to create intense molecular beams of large molecules and to ingenious
multi-resonance laser spectroscopy schemes that allow to disentangle the signatures of different
isomers~\cite{Suenram:JACS102:7180} present even in cold molecular beams~\cite{Rizzo:JCP83:4819}.
However, for many novel studies it would be very advantageous, or even necessary, to separate the
individual isomers, to select quantum states, or to slow down these molecules. There is a large
interest in performing coherent x-ray diffractive imaging of
biomolecules~\cite{Neutze:Nature406:752} using novel free-electron-laser x-ray sources. It would be
very advantageous to perform initial experiments on well-defined targets: ensembles of molecules all
with the same structure (conformation) and all strongly oriented in space. Similar arguments can be
made for high-harmonic generation~\cite{Itatani:PRL94:} or tomographic orbital
reconstruction~\cite{Itatani:Nature432:867} experiments using (large) molecules.

In this paper we will describe the different experimentally proven methods for the manipulation of
the motion of large molecules. We will focus on the manipulation of the translational motion, where
we will present detailed descriptions and experimental results obtained in our laboratory. We place
special emphasis on the ability to separate conformers with these experiments and will compare the
approaches with one another.

\section{Experimental approaches}
\label{sec:experiment}

Several complementary experimental approaches for the manipulation of the translation and for the
state selection of large neutral molecules exist. Here we limit the discussion to the use of
inhomogeneous electric fields, although similar experiments could be performed using
magnetic~\cite{Gerlach:ZP9:349, Vanhaecke:PRA75:031402, Narevicius:PRL100:093003} or laser
fields~\cite{Zhao:PRL85:2705, Fulton:PRL93:243004}. All these methods rely on the strong cooling
provided by supersonic expansions, where rotational and translational temperatures in the moving
frame of the molecular beam on the order of \unit[1]{K} are routinely
achieved~\cite{_Scoles:MolBeam:1}.

Whereas this article describes the manipulation of the translational motion of molecules, methods to
manipulate the rotational motion have also been demonstrated. There are several electric field based
methods that are applicable for large molecules, which are generally asymmetric rotors. The
conceptionally simplest method to confine the angular distribution of polar molecules is the
interaction of the molecular dipole with a strong homogeneous electric field, as proposed
independently by Loesch \etal~\cite{Loesch:JCP93:4779} and by Friedrich and
Herschbach~\cite{Friedrich:Nature353:412}. This approach has been experimentally demonstrated many
times and is summarised elsewhere~\cite{Loesch:ARPC46:555, Kong:IJMPB15:3471}. This ``brute force
orientation'' of large molecules has been exploited, for example, to determine transition moment
angles in the molecular frame~\cite{Castle:JCP112:10156, Dong:Science298:1227}. Applying strong,
non-resonant laser fields to the molecules also provokes angular
confinement~\cite{Stapelfeldt:RMP75:543}. The crucial influence of the population of rotational
states has been experimentally investigated~\cite{Kumarappan:JCP125:194309}. Clearly, the
state-selection of the lowest rotational states, performed by the experiments described below,
allows for considerably stronger degrees of alignment~\cite{Holmegaard:PRL102:023001}. Recently,
strong alignment and orientation by mixed dc electric and laser fields~\cite{Friedrich:JCP111:6157,
   Friedrich:JPCA103:10280} has been demonstrated for the large asymmetric top molecule
iodobenzene~\cite{Holmegaard:PRL102:023001}. For a more-in-depth discussion of alignment and
orientation the reader is referred to the existing excellent reviews~\cite{Stapelfeldt:RMP75:543,
   Seideman:AAMOP52:289}

\subsection{Electric beam deflection}

\subsubsection{General description}

The possibility to deflect polar molecules from a molecular beam using inhomogeneous electric fields
was first described by Kallmann and Reiche in 1921.\cite{Kallmann:ZP6:352}$^,$\footnote{Interesting
   enough these studies were performed at the \emph{Kaiser Wilhelm Institut für physikalische Chemie
      und Elektrochemie}, the predecessor of the Fritz Haber Institut in Berlin.} Already in 1926
Stern suggested that the technique could be used for the quantum state separation of small diatomic
molecules at low temperatures~\cite{Stern:ZP39:751}. The electric deflection of a molecular beam was
experimentally demonstrated by Wrede -- a doctoral student of Otto Stern -- in 1927 and the dipole
moment of KI was determined~\cite{Wrede:ZP44:261}. In 1939 Rabi introduced the molecular beam
resonance method, by using two deflection elements of oppositely directed gradients in succession to
study the quantum structure of atoms and molecules~\cite{Rabi:PR55:526}. Already by 1956 a number of
different approaches to beam deflection were discussed in Ramsey's
textbook\cite{Ramsey:MolBeam:1956}. Since then, electric beam deflection has been used extensively,
for example, to determine polarisabilities and dipole moments of clusters and
molecules~\cite{Moro:Science300:1265, Broyer:PhysScr76:C135}.

\begin{figure}
   \centering%
   \includegraphics[width=\figwidth]{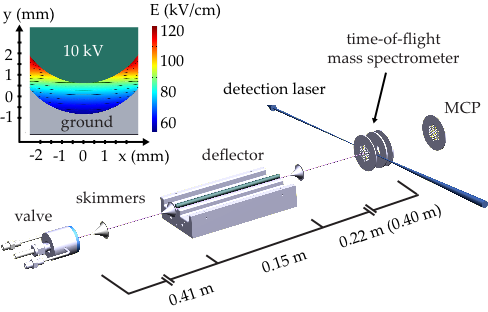}%
   \caption{(Colour online): Experimental setup of the electric $m/\mu$ deflector for quantum-state
      selection of large molecules. An internally cold molecular beam is produced by expanding a
      mixture of a few percent of the investigated molecule in several bar of rare gas. The
      resulting supersonic jet is skimmed a few centimetres downstream the nozzle for differential
      pumping and again about 41~cm downstream the nozzle by a 1~mm diameter skimmer for beam
      collimation. Then the molecular beam enters the electric deflector, where the inhomogeneous
      field shown enlarged in the inset provides a force on polar molecules in the vertical
      direction. Behind the deflector another 1.5~mm skimmer provides further differential pumping
      for the molecules entering the detection chamber, where the (vertical) deflection profiles can
      be measured by scanning a pulsed dye laser up and down in an resonance-enhanced
      multi-photon-ionisation time-of-flight mass-spectrometry detection scheme.}
   \label{fig:setup:deflector}%
\end{figure}

Commonly, the so-called two-wire-field geometry, depicted in figure~\ref{fig:setup:deflector}, is
used~\cite{Ramsey:MolBeam:1956}, but more advanced electrode geometries have also been employed, for
example, in recent matter-wave interferometry experiments~\cite{Berninger:PRA76:013607}.
Nevertheless, the classical two-wire field is widely employed, due to its experimental simplicity
and the quite good results that can be obtained with it.

\subsubsection{Deflection of large molecules}

Using an experimental setup as shown in figure~\ref{fig:setup:deflector} we have, in collaboration
with the group of Henrik Stapelfeldt (University of Aarhus, Denmark), deflected large molecules,
\ie, iodobenzene (C$_6$H$_5$I), and performed laser alignment and mixed-field orientation of the
produced state-selected samples~\cite{Holmegaard:PRL102:023001}. For these studies it was crucial to
already start with cold molecular beams as they are produced in high-stagnation-pressure
expansions~\cite{Hillenkamp:JCP118:8699}.
\begin{figure}
   \centering%
   \includegraphics[width=\figwidth]{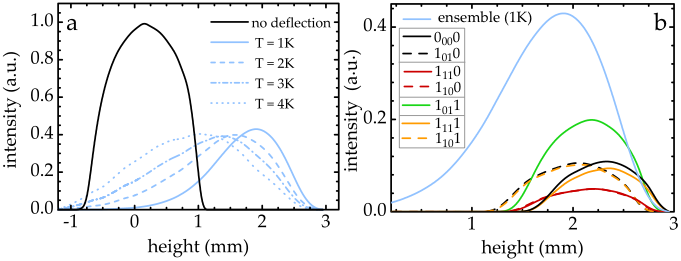}%
   \caption{(Colour online): Deflection profiles a) for molecular beams of benzonitrile at different
      rotational temperatures and b) for a molecular beam of benzonitrile at 1~K and individual
      profiles of the lowest energy rotational states. The intensities of all individual states are
      according to their population at 1~K and are scaled by a factor 10. All calculations are made
      for the setup shown in figure~\ref{fig:setup:deflector} using a distance from deflector to
      detector of \unit[22]{cm}.}
   \label{fig:deflector:temperature}%
\end{figure}
In figure~\ref{fig:deflector:temperature}\,a the calculated deflection profiles of benzonitrile from
molecular beams of varying temperature is shown. For the lowest temperature (1~K) almost the
complete beam can be deflected. With increasing temperature, the overall deflection significantly
decreases and instead a considerable broadening of the beam is obtained. Due to the large number of
quantum states involved we have not performed quantitative calculations for temperatures above 4~K,
but approximate calculations suggest that already at 10 K no significant deflection is obtained
anymore. In figure~\ref{fig:deflector:temperature}\,b the contribution of the lowest individual
rotational states, \ie, all states with $J=0$ or $1$, to the deflection profiles is depicted. These
states belong to four different Hamiltonian matrix blocks as shown in the figure legend (see
appendix~\ref{sec:stark-calc} for details) and the lowest energy states, shown as solid lines in
figure~\ref{fig:deflector:temperature}\,b, are all very polar and behave very similar under the
influence of the electric field. Especially the states $J_{K_aK_c}=0_{00}$ and $1_{11}$ behave
extremely similar -- they are the ground states of para- and ortho-benzonitrile, respectively -- and
differ practically only in intensity, which is due to the difference in population.

In any case, using a focused laser or a narrow slit one could perform experiments selectively with
the molecular ensemble at a given height. Choosing, for example, a height of $z=\unit[2.7]{mm}$ at
\unit[1]{K}, where the density is \unit[5]{\%} of the density at the peak of the free flight, only a
few quantum states are populated, which are all very polar and have very similar effective dipole
moments $\mueff$ (the negative gradient of the Stark curve). Therefore, such an ensemble can be
strongly oriented using dc electric fields or mixed dc and laser fields.

For cold beams of small molecules with large rotational constants and, therefore, only a few
rotational states populated in the molecular beam, this method would allow the preparation of
samples of individual rotational states~\cite{Stern:ZP39:751}. For low-field-seeking states this can
also be achieved using electric multipoles, which also provide transverse
focusing~\cite{Reuss:StateSelection, Stolte:StateSelectedScattering:1988}, or, even cleaner, using
the Stark decelerator~\cite{Bethlem:PRL83:1558, Meerakker:FD142}. The deflector, nevertheless, also
allows to individually address high-field-seeking states, \eg, absolute ground states.

The large differences between the calculated deflection profiles for benzonitrile at different
temperatures also demonstrate that deflection profiles can be used as a sensitive measure of the
internal temperatures of molecular beams~\cite{Moro:PRA75:013415}, especially for low rotational
temperatures, where the strongly polar quantum states are populated the most and where meaningful
quantum-mechanical calculations can still be performed.

\subsubsection{Conformer selection}
\label{sec:deflector:conformer-selection}

For more complex molecules typically multiple conformers are present in a molecular
beam~\cite{Rizzo:JCP83:4819}. These conformers all have the same mass, but typically exhibit largely
different dipole moments. Therefore, one can use the distinct forces exerted on the molecules by
inhomogeneous electric fields to spatially separate the molecules based on their $m/\mu$ ratios. We
have already demonstrated this for 3-aminophenol using dynamic focusing, see
section~\ref{sec:selector} and reference\hspace{1ex} \onlinecite{Filsinger:PRL100:133003}. Here, we
want to discuss the possibilities to use an electric deflector for the same purpose.

\begin{figure}
   \centering
   \includegraphics[width=0.66\figwidth]{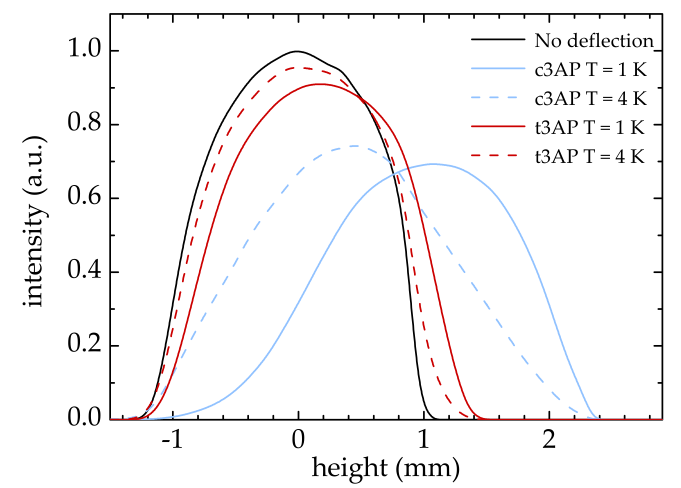}%
   \caption{(Colour online): Simulated deflection profiles of a beam of cis-3-aminophenol (c3AP) and
      trans-3-aminophenol (t3AP) using the setup described above (with a deflector-to-detector
      distance of \unit[40]{cm}) at a voltage of \unit[12]{kV} applied to the rod and rotational
      temperatures of 1 and \unit[4]{K}. In these simulations the populations of cis- and
      trans-3-aminophenol in the original beam are assumed to be equal.}
   \label{fig:deflection:3AP}
\end{figure}

In figure~\ref{fig:deflection:3AP} the simulated deflection profiles of the cis and trans conformers
of 3-aminophenol are shown for the experimental setup described above and the known molecular
parameters~\cite{Filsinger:PCCP10:666}.
For clarity we have assumed equal population of the two species in these simulations; the actual
populations can be estimated to be 1:4 for cis-:trans-3-aminophenol from their intensities in
electronic spectra. From the simulations at 1~K it is clear that a large fraction of the
cis-3-aminophenol conformers can be deflected out of the original beam and even out of the
distribution of the trans conformer. Moreover, even at 4~K one can foresee to perform experiments
with a pure sample of cis-3-aminophenol. On the other hand, no clean sample of the less polar
trans-3-aminophenol conformer can be produced this way. This is even more so, since typically the
internal temperatures of the supersonic beams are not thermal~\cite{WU:JCP91:5278} and a
high-temperature fraction, corresponding to mostly unpolar quantum states, exists. Therefore, in
general only one, the most polar conformer, can be separated from the others. These results should
be compared to the demonstrated selection of the conformers of 3-aminophenol using the AG focusing
selector described in section~\ref{sec:selector}, where each conformer can be addressed
individually.

\subsection{Dynamic focusing selectors}
\label{sec:selector}

\subsubsection{General description}

Whereas deflection experiments allow the spatial dispersion of quantum states, they do not provide
any focusing. For small molecules in low-field-seeking states this issue could be resolved using
multipole focusers with static electric fields, which were developed independently in
Bonn~\cite{Bennewitz:ZP139:489, Bennewitz:ZP141:6} and in New York~\cite{Gordon:PR95:282,
   Gordon:PR99:1264} in 1954/55. About ten years later, molecular samples in a single rotational
state were used for state specific inelastic scattering experiments by the Bonn
group~\cite{Bennewitz:ZP177:84} and, shortly thereafter, for reactive
scattering~\cite{Brooks:JCP45:3449, Beuhler:JACS88:5331}.

However, for large molecules all quantum states are practically hfs at the relevant electric field
strengths. Therefore, focusing with static electric fields is not possible. Instead, one has to
retreat to dynamic focusing schemes, which are also used in the operation of the LINAC, quadrupole
ion guides, or Paul traps for charged particles. Dynamic focusing of neutral polar molecules was
described by Auerbach \etal\ in 1966~\cite{Auerbach:JCP45:2160}. Successively, it was experimentally
demonstrated~\cite{Kakati:PLA24:676, Kakati:PLA28:786, Gunther:ZPC80:155, Lubbert:JCP69:5174} and
successfully applied in maser experiments~\cite{Kakati:JPE4:269, Laine:Entropie42:165}.

\begin{figure}
   \centering%
   \includegraphics[width=\figwidth]{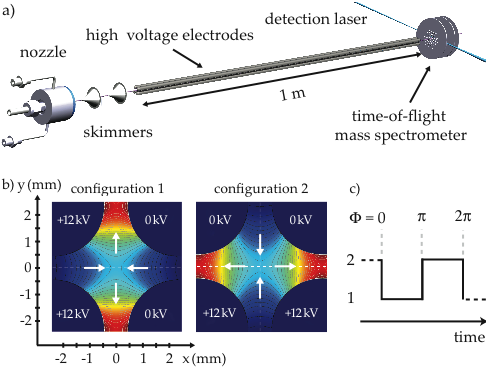}%
   \caption{(Colour online): a) Experimental setup of the dynamic-focusing electric $m/\mu$ selector
      for quantum-state selection of large molecules. An internally cold molecular beam is produced
      by expanding a mixture of a few percent of the investigated molecule in several bar of rare
      gas. The resulting supersonic jet is skimmed a few centimetres downstream from the nozzle for
      differential pumping and again about \unit[18]{cm} downstream from the nozzle by a
      \unit[xs1]{mm} diameter skimmer for beam collimation. Then the molecular beam enters the
      electric selector, where the switched inhomogeneous fields given in the inset (b) provide a
      focusing force on polar molecules towards the molecular beam axis. The transmission through
      the selector is monitored in a resonance-enhanced multi-photon-ionisation time-of-flight
      mass-spectrometry detection scheme. c) Definition of the phase of the switching cycle, where
      configuration~$1$ corresponds to focusing in $x$-direction and $2$ to focusing in
      $y$-direction for molecules in high-field-seeking states.}
   \label{fig:setup:selector}%
\end{figure}

In these early experiments the switching frequency was defined by the beam velocity and the
electrode geometry, requiring a setup with alternatingly oriented electric field lenses. Nowadays,
however, it is possible to electronically switch the necessary electric fields and one can use, for
example, a four-wire setup, shown in figure~\ref{fig:setup:selector}, with varying voltages applied
to create the necessary alternating gradient focusing at any frequency and always applying the
maximum field strength and gradient, \ie, always applying maximal forces.

\subsubsection{Focusing of molecules in low- and high-field-seeking states}

\begin{figure}
   \centering%
   \includegraphics[width=0.66\figwidth]{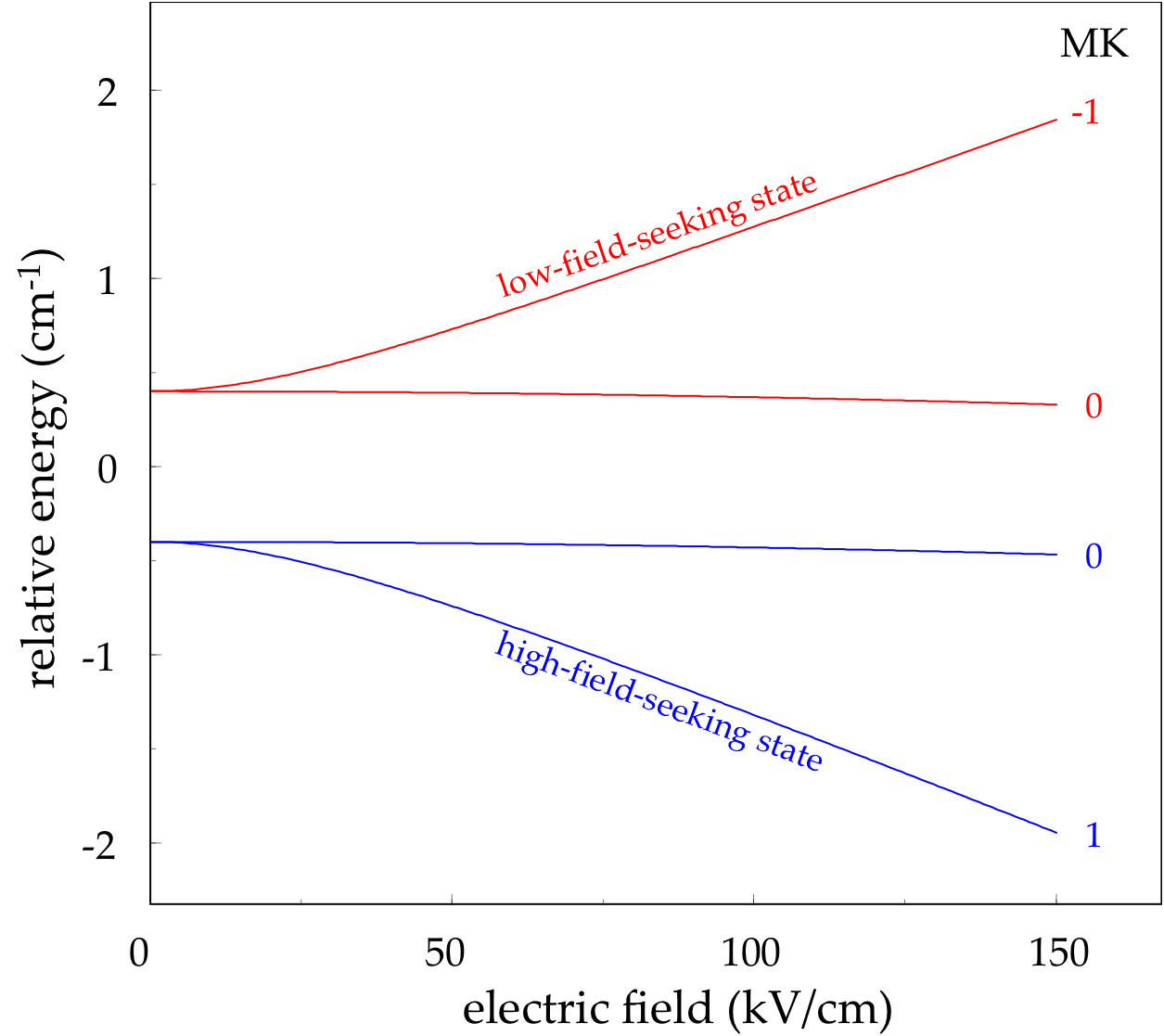}%
   \caption{(Colour online): The energy of ammonia (NH$_3$) in its $J_K=1_1$ rotational state as a
      function of electric field strength (neglecting hyperfine structure).}
   \label{fig:selection:ammonia:stark}%
\end{figure}

\begin{figure}
   \centering%
   \includegraphics[width=\figwidth]{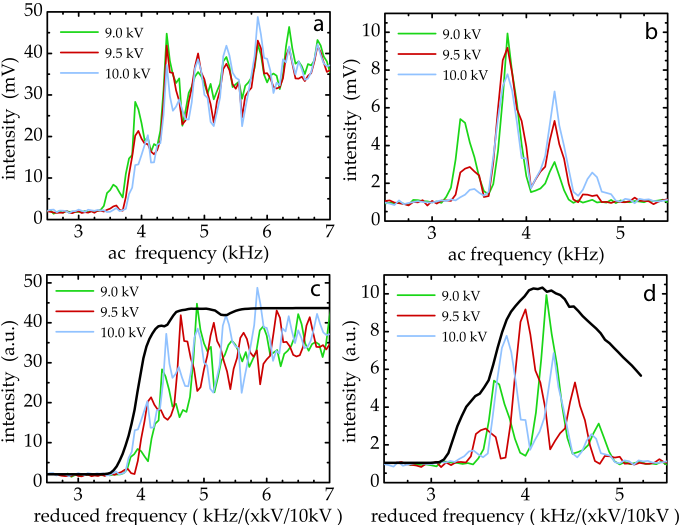}%
   \caption{(Colour online): Measured transmission through the $m/\mu$ selector for ammonia in its
      lfs ($MK=-1$, left graphs) and hfs ($MK=1$, right graphs) states of the $J_K=1_1$ rotational
      state. In the upper plots the experimental transmission as a function of the switching
      frequency between the two electric field configurations, shown in
      figure~\ref{fig:setup:selector}\,b, is plotted for different applied voltages. The observed
      strong modulation in all measurements is due to changes in the end-phase of the switching
      cycle and, therefore, the overlap between the focused detection laser and the molecular
      packet. The differences in the peak intensities is due to the changed frequency dependence
      according to the focusing strength at the different voltages. In the lower graphs the same
      data is plotted as a function of a reduced frequency $\tilde{f}=f/(U/10^4~\text{V})$ with the
      applied switching frequency $f$ and voltage $U$. The envelope of these data nicely represents
      the simulated overall transmission through the selector (solid black lines), independent of
      the detection (overlap) function.}
   \label{fig:selection:ammonia}%
\end{figure}

\begin{figure}
   \centering
   \includegraphics[width=\figwidth]{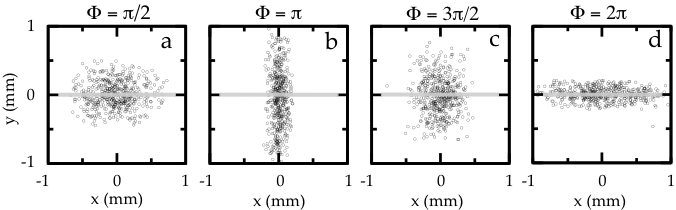}%
   \caption{Transverse phase-space distribution of the molecular packet in the detection region as a
      function of the end-phase $\Phi$ of the switching cycle depicted in
      figure~\ref{fig:setup:selector}\,c. The horizontal grey lines depict the area probed by the
      focused detection laser.}
   \label{fig:selection:phasespace}
\end{figure}

In order to characterise the operation of the AG focuser we have performed initial experiments using
ammonia (NH$_3$) in its $J_K=1_1$ rotational state. NH$_3$ in this state, the rotational ground
state of para ammonia, exhibits a quadratic Stark effect at low and moderate electric field
strengths that converges to a linear Stark shift once the Stark energy is comparable to the
inversion splitting, as shown in figure~\ref{fig:selection:ammonia:stark}.
Ammonia molecules in the lfs $MK=-1$ component can be focused using a static quadrupole field. This
was already exploited by Gordon and Townes in the original demonstration of the
MASER~\cite{Gordon:PR99:1264} and was performed by us for initial optimisation of expansion
conditions and laser detection. However, ammonia molecules in both polar quantum states ($MK=-1$ and
$+1$) can be confined to the beam axis using AG focusing. This is demonstrated by the measurements
in figure~\ref{fig:selection:ammonia}.
Here, the transmissions of NH$_3$ in its $J_K=1_1,MK=-1$ and $J_K=1_1,MK=1$ states are plotted as a
function of the frequency used to switch between the two electric field configurations. The
individual states are selectively detected by choosing appropriate 2+1-REMPI transitions. The
transmission for the molecules in hfs quantum states (figure~\ref{fig:selection:ammonia}\,b) shows a
frequency dependence as expected: At low switching frequencies the molecules are strongly defocused
in one direction and lost from the focuser. At high frequency the time averaged potential is
approximately flat and no focusing occurs, therefore, the transmission is also low. In between, at
the appropriate switching frequency, AG focusing works and the transmission is high. The
experimentally measured transmission is strongly modulated in that frequency range. This modulation
is due to the overlap between the strongly focused detection laser ($w_L\approx\unit[40]{\mu{}m}$)
and the changing shape of the molecular packet. This shape strongly depends on the phase in the
switching cycle as depicted in figure~\ref{fig:selection:phasespace}.
Due to the short focusing device -- corresponding to a short residence time of the molecules in the
device -- the start phase, the end phase, and the switching frequency cannot independently be
optimised. We chose to optimise the start phase for optimum transmission and then the end phase is
determined from the applied switching frequency. In order to reduce these experimental artifacts, we
measured the transmission curves for different applied high-voltages $U$. For comparison the
frequencies $f$ of these different measurements are converted to a reduced frequency
$\tilde{f}=f/(U/10^4~\text{V})$ taking into account the effective focusing strength. The resulting
transmission characteristics are displayed in figure~\ref{fig:selection:ammonia}\,d. The envelope of
these measurements clearly represents the expected overall transmission curve, free of strong
effects due to changes in the detection efficiency.

The measurements on ammonia in its lfs $J_K=1_1$, $MK=-1$ states, shown in
figure~\ref{fig:selection:ammonia}\,a and c, demonstrate the versatility of the device. Clear
evidence of dynamic focusing is obtained, with similar characteristics as for the hfs states. Two
differences in the frequency dependence of the transmission are obvious: at high frequencies the
transmission does practically not decrease, and the low-frequency-onset of the transmission curve is
shifted towards higher frequencies. The latter demonstrate the quantum state dependence of the
process and that, at least in principle, one could use the device to separate the two quantum
states. Moreover, since focusing and defocusing forces are interchanged between the lfs and hfs
states, the phase effects shown in figure~\ref{fig:selection:phasespace} are shifted by $\pi$, which
can also be used to discriminate between the two states in laser experiments or by narrow slits in
the beam path.

The large transmission at high frequencies is due to the fact that the time-averaged potential (the
average of the two potentials in figure~\ref{fig:setup:selector}) is not really flat, but has a
minimum of electric field on the molecular beam axis. Therefore, under these conditions molecules in
lfs states are focused, whereas molecules in hfs states are defocused. This minimum has a depth of
\unit[7.5]{kV/cm}, corresponding to a 2-dimensional trap depth for ammonia in its lfs state of
\unit[0.11]{K}~\footnote{A similar field could, of course, be created by applying dc voltages in a
   quadrupole arrangement of $\pm 0.6$~kV, although the trap depth and characteristics would be
   different due to the approximately quadratic Stark effect of NH$_3$ at the resulting low field
   strengths.}.

\subsubsection{Conformer selection}
\label{sec:selector:conformer-selection}

For given applied voltages the transmission through the selector is a function of the effective
dipole moment \mueff{} of the molecule's quantum state and of the switching frequency
$f$~\cite{Auerbach:JCP45:2160, Bethlem:JPB39:R263}: the larger the dipole moment, the higher the
optimal switching frequency. Consequently, for a given switching frequency $f$, the transmission
depends on the molecules \mueff. This can, in principle, be exploited for the separation of
individual quantum states. As the dipole moments of different conformers of large molecules often
vary quite dramatically, one can use the selective focusing to separate the individual conformers
from each other. This has been demonstrated for the cis- and trans-conformers of
3-aminophenol~\cite{Filsinger:PRL100:133003}.

In comparison to the separation by deflection proposed in
section~\ref{sec:deflector:conformer-selection}, the AG focusing selector can provide enhanced
transmission for any polar conformer. However, the focused molecules are confined to the molecular
beam axis, where there is a background of molecules in unpolar quantum states and of atomic seed
gas. In principle, the background could be removed by a beam stop on the molecular beam
axis~\cite{Reuss:StateSelection}. However, such a beam stop would take away a very large fraction of
the beam intensity. Moreover, the central part of the beam is typically also the internally coldest,
and, therefore, the most polar one. A similar effect can be obtained by tilting the focuser against
the incoming molecular beam axis. We have done this in the selection experiments on 3-aminophenol
and it did provide a somewhat improved contrast, although even then a considerable amount of
background was observed~\cite{Filsinger:PRL100:133003}.

\subsubsection{Resolution}
\label{sec:selector:resolution}

\begin{figure}
   \centering%
   \includegraphics[width=\figwidth]{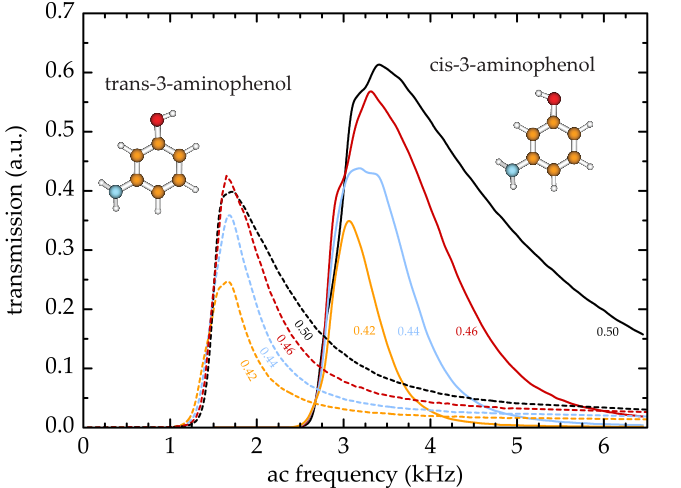}%
   \caption{(Colour online): Calculated transmission through the selector for cis- (full lines) and
      trans-3-aminophenol (dashed lines) in their rotational ground state for different duty cycles.
      The duty cycle for each individual simulation is given in the figure.}
   \label{fig:selection:resolution}%
\end{figure}

In the conformer selection experiment described above, the focusing selector has been operated under
conditions of maximum throughput, the equivalent of a quadrupole \emph{ion guide}. Just like the
$m/q$-resolution of quadrupole mass spectrometers can be improved at the cost of transmission by
applying a dc offset, the $m/\mu$-resolution of the focusing selector can be improved at the cost of
transmission by changing the duty cycle of the applied high-voltage switching sequence. This effect
is demonstrated by the simulations in figure~\ref{fig:selection:resolution}, where the transmission
as a function of frequency is shown for the ground states of 3-aminophenol for duty cycles
$\tau_x/(\tau_x+\tau_y)=0.5$, $0.46$, $0.44$, and $0.42$, respectively.
Here $\tau_x$ and $\tau_y$ are the fractions of a switching period for which $x$- and $y$-focusing
are applied (with $\tau_x+\tau_y=1$). From the calculations it is clear that the changed duty cycle
improves the resolution, whereas the intensity decreases simultaneously. In principle, the effect is
completely equivalent to the dc offset in a quadrupole guide for charged particles. However, it has
to be taken into account that the charged particles experience a harmonic potential inside the
guide, whereas the large neutral molecules, discussed here, experience a strongly non-harmonic
force. This is due to the often highly non-linear Stark effect (see
figure~\ref{fig:BN:Stark-curves}) and the higher-order terms in the created electric
fields~\cite{Bethlem:JPB39:R263, Tarbutt:NJP10:073011}. Nevertheless, the effect can be used to
separate species with smaller dipole moment differences or, for the same samples, to achieve
stronger discrimination between different species. We are currently experimentally verifying these
simulations.

It has also to be taken into account, that the molecules experience only a few electric field
switching periods inside the current device. Together with the initial spatial spread of the
molecular packet, this results in the switching frequency not being well-defined for the molecular
ensemble.

In order to be able to routinely operate under conditions with higher resolution we plan to set up a
focuser with an improved overall transmission, by scaling up the transverse phase-space acceptance,
and a larger residence time, by making the device longer. This should enable us to separate
individual conformers of many complex molecules.

\subsection{Alternating gradient decelerator}

\begin{figure}
   \centering%
   \includegraphics[width=\figwidth]{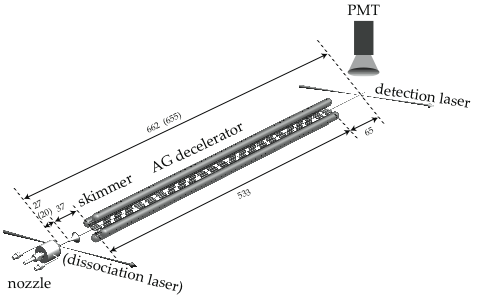}%
   \caption{Experimental setup of the alternating-gradient decelerator. An internally cold molecular
      beam is produced by expanding a mixture of a few percent of the investigated molecule in
      approximately 1~bar of rare gas. In experiments with OH, an expansion of HNO$_3$ is irradiated
      by a pulsed laser inside the expansion region in order to photodissociate HNO$_3$ to form OH
      which is successively cooled in the remaining expansion. The resulting supersonic jet is
      skimmed a few centimetres downstream from the nozzle for differential pumping and enters the
      54~cm long decelerator. The time-resolved transmission of individual quantum states through
      the decelerator is monitored in a laser-induced-fluorescence detection scheme using a
      narrow-linewidth continuous-wave dye laser~\cite{Wohlfart:PRA77:031404,
         Wohlfart:PRA78:033421}.}%
   \label{fig:AG-dec:setup}%
\end{figure}

The alternating gradient decelerator, depicted in figure~\ref{fig:AG-dec:setup}, combines the
dynamic focusing of the selector with deceleration equivalent to the Stark
decelerator~\cite{Meerakker:NatPhys4:595} and it has been described in detail
elsewhere~\cite{Auerbach:JCP45:2160, Bethlem:JPB39:R263}.
Generally, alternating-gradient deceleration is applicable to molecules in any polar quantum state,
\ie, it allows the deceleration of molecules in lfs states and in hfs states. This has been
demonstrated for the deceleration of OH radicals in their lfs and hfs components of the \ohstate{}
$\Lambda$-doublet~\cite{Wohlfart:PRA78:033421}. Experiments on the alternating-gradient deceleration
of diatomic molecules in hfs states have also been performed for the diatomic molecules
CO*~\cite{Bethlem:PRL88:133003} and YbF~\cite{Tarbutt:PRL92:173002}.

\begin{figure}
   \centering%
   \includegraphics[width=0.5\figwidth]{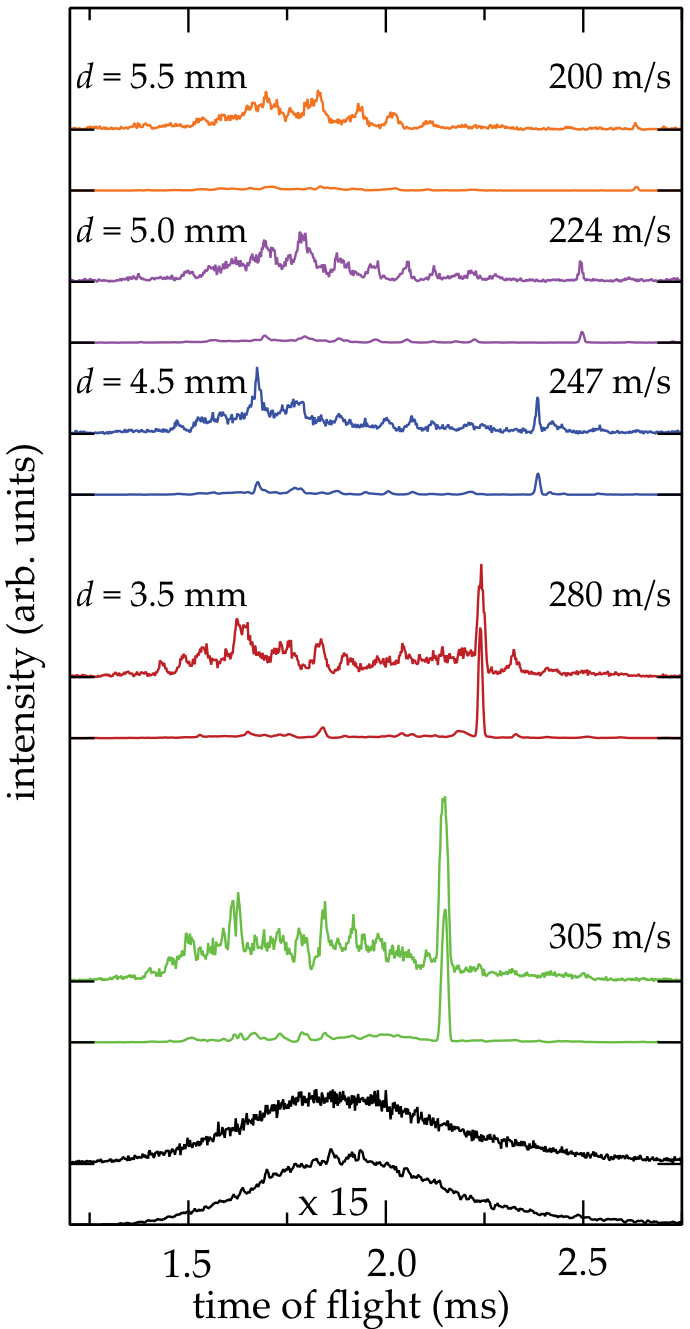}%
   \caption{AG focusing and deceleration of OH in its lfs state from 305~m/s to varying velocities
      as specified next to the individual measurements. For a deceleration strength of $d=5.5$
      molecular packets with a centre velocity of 200 m/s are obtained.}
   \label{fig:AG-dec:OH-200}%
\end{figure}

It has also been laid out that, in order to decelerate molecules in lfs states using the AG
decelerator, the fields have to be switched on twice per electrode pair -- once in between
successive pairs in order to provide longitudinal bunching and deceleration, and once inside the
electrode pair, in order to provide transverse focusing. In figure~\ref{fig:AG-dec:OH-200} a
deceleration sequence of OH radicals in their lfs state from 305 to \unit[200]{m/s} using 27 AG
lenses is given.
This is the lowest-velocity molecular packet obtained from an AG decelerator so far and it clearly
demonstrates the versatility of the AG decelerator. It has to be taken into account, however, that
the phase-space acceptance is at least one order of magnitude smaller than for deceleration of OH
radicals in their lfs state using the normal Stark decelerator~\cite{Wohlfart:PRA78:033421}.

For molecules in hfs states the fields are switched on when the molecules are inside the electrode
pair (AG lens), where the molecules are focused in transverse direction. When the molecules exit the
lens, they are decelerated before the field is switched off again sometime before the molecules
reach the minimum of the electric field which is at the centre between two successive lenses. In
principle, the focusing works the same way as described for the selector. However, in the
decelerator the switching frequency is determined by the distance of successive electrode pairs and
the velocity of the molecules and can, therefore, not be varied independently. Moreover, in order to
obtain maximal fields on the molecular beam axis for the deceleration process, the fields are
created by only two cylindrical electrodes and their geometric orientation defines the focusing and
defocusing direction. However, one can change the overall focusing strength for the given
geometry/directions by the duration the fields are switched on, or equivalently, by the distance $f$
the molecules travel inside the electrodes. In principle the focusing strength can also be changed
by changing the applied voltage. However, this would reduce the maximum field strength and,
therefore, the deceleration strength, and it can also not as quickly be adjusted during the
experiment.

\subsubsection{Alternating gradient deceleration of large molecules}

\begin{figure}[t]
   \centering%
   \includegraphics[width=\figwidth]{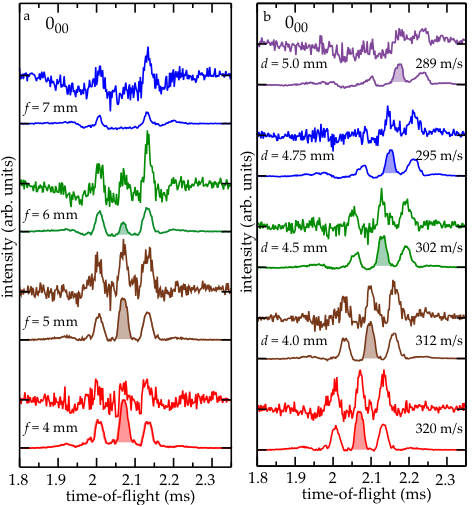}%
   \caption{(Colour online): Alternating-gradient focusing and deceleration of benzonitrile from a
      molecular beam of 320~m/s~\cite{Wohlfart:PRA77:031404}. In the left column no overall
      deceleration is performed and the effects of a changing focusing strength are demonstrated. In
      the right column packets of benzonitrile molecules are decelerated from 320~m/s to
      successively lower final velocities as indicated in the figure.}
   \label{fig:AG-dec:BN}%
\end{figure}

The experimental setup of the alternating gradient decelerator is shown in
figure~\ref{fig:AG-dec:setup}. Using this setup we have AG focused and decelerated benzonitrile in
its absolute ground state, and the obtained difference-time-of-flight profiles (see
reference\hspace{1ex}\ \onlinecite{Wohlfart:PRA77:031404} for details) are shown in
figure~\ref{fig:AG-dec:BN}.
Figure~\ref{fig:AG-dec:BN}\,a illustrates the focusing behavior of the $0_{00}$ ground state of
benzonitrile for a constant velocity of the synchronous molecule of 320~m/s. The high voltages are
switched on symmetrically around the centers of the AG lenses. Therefore, the molecular packet is
focused in both transverse directions as well as in the longitudinal direction (bunching), but no
change of the synchronous velocity occurs. In the experiments three packets of focused molecules are
observed. The central peaks of the TOF distributions occur 2.07~ms after the molecules exit the
nozzle. These packets contain the synchronous molecule. Hereafter, they are referred to as the
``synchronous packets'', and they are shaded in the simulated TOF distributions. The peaks at
earlier and later arrival times correspond to molecular packets leading and trailing the synchronous
packet by one AG lens (or 20~mm), respectively. These packets are also focused in all three
dimensions. However, due to the lens pattern in our setup, they experience only 2/3 of the lenses at
high voltage. This results in a reduced overall focusing for these packets.

For the focusing of the $0_{00}$ state it is seen that the synchronous packet is most intense for a
focusing length of $f=5$~mm. Under these conditions approximately $10^5$ molecules per quantum state
per pulse are confined in the phase-stable central peak, corresponding to a density of
$\unit[10^8]{cm^{-3}}$. For smaller focusing lengths a shallower time-averaged confinement potential
is created, and less molecules are guided through the decelerator. For larger focusing lengths the
molecular packet is over-focused, also resulting in a decreased transmission. For
\unit[\hbox{$f=7$}]{mm} the over-focusing is so severe that the synchronous packet completely
disappears. As expected, the non-synchronous packets benefit from the increased focusing lengths.

Figure~\ref{fig:AG-dec:BN}\,b presents the results for the deceleration of benzonitrile in its
$0_{00}$ state. The bottommost (red) trace shows focusing experiments for a constant velocity of the
synchronous molecule of 320~m/s using the optimum focusing lengths. All other traces show
experiments in which the synchronous packet is decelerated from 320~m/s to successively lower
velocities, resulting in later arrival times at the detector. In these experiments the deceleration
strength $d$ is given as the position on the molecular beam axis behind the centre of the AG lens,
where the electric field is switched off. Using $d=5$~mm, the packet is decelerated to 289~m/s,
corresponding to a reduction of the kinetic energy by 18~\%. The observed intensities of the
non-synchronous packets decrease faster upon increasing $d$. Because the molecules in these packets
miss every third deceleration stage, their trajectories are not stable and they are only observed
due to the finite length of the decelerator. When deceleration to lower velocities is performed by
increasing $d$ the signal intensity decreases due to the reduction in phase-space acceptance.
However, one could also decelerate to lower velocities by increasing the number of AG lenses for a
given value of $d$. In this case, in principle, no decrease in the intensity of the synchronous
packet is expected due to the phase stability of the deceleration process~\cite{Bethlem:JPB39:R263}.
For all deceleration measurements, the relative intensities and the arrival times of the molecular
packets at the detector are nicely reproduced by the trajectory simulations.

\subsubsection{Towards the trapping of large molecules}
\label{sec:discussion:trapping}

Once large molecules in polar quantum states are decelerated down to slow velocities, \ie, below
approximately \unit[25]{m/s} for molecules like benzonitrile or 3-aminophenol, they can be confined
in ac electric traps. Such traps have already been demonstrated for slow ammonia molecules in hfs
states~\cite{Veldhoven:PRL94:083001, Bethlem:PRA74:063403, Schnell:JPCA111:7411}, which were
produced from Stark decelerated ammonia in lfs states \emph{via} a microwave
transition~\cite{Veldhoven:EPJD31:337}. Trapping lifetimes will be limited by collisions with
background gas, by blackbody radiation~\cite{Hoekstra:PRL98:133001}, and by non-adiabatic following
of Stark curves. Even though the collisional cross sections for losses due to background gas will be
different for large molecules compared to the ones for small molecules, the effects should be small.
Nonadiabatic dynamics on the Stark potential energy curves do not occur for the lowest quantum
state. For higher states it can be avoided, at least for low-lying rotational states, by providing
finite minimum electric field strength on the order of \unit[10]{kV/cm}. This also efficiently
prohibits Majorana transitions. Since ac traps must provide a saddle point in the trap centre this
requirement is practically automatically fulfilled.

The blackbody-radiation lifetimes of individual states can be calculated from the radiation density
and the possible transitions. For the rovibronic ground state ($\JKaKc=0_{00}$) two types of
transitions have to be taken into account: rotational excitation into the rotational excited states
that can be reached by an allowed transition and vibrational excitations. We have performed the
calculation for the lifetime of benzonitrile in its rovibronic ground state as an example of a
prototypical large polar molecule. At \unit[300]{K} the rotational transition rate for the
$1_{01}\leftarrow0_{00}$ rotational $a$-type transition at
\unit[2.76]{GHz}~\cite{Wohlfart:JMolSpec247:119} is $\unit[10^{-5}]{Hz}$. In order to estimate the
vibrational transition rate, we have performed a geometry optimisation and normal coordinate
analysis at the B3LYP/aug-cc-pVTZ level of theory using Gaussian 2003~\cite{Gaussian:2003C02}, which
yields transition strengths for all 33 fundamental transitions. This results in an overall rate for
loss of ground state molecules due to blackbody radiation at room temperature (\unit[300]{K}) of
less than \unit[1/4]{Hz}. The resulting trapping lifetime of \unit[4]{s} is larger than currently
achieved trapping times of small molecules in ac traps~\cite{Veldhoven:PRL94:083001,
   Veldhoven:thesis:2006, Bethlem:PRA74:063403, Schnell:JPCA111:7411} and does not pose a major
restriction on the trapping of such large molecules. However, the major loss rate is due to
low-frequency vibrations with transition frequencies between 400 and $\unit[800]{cm^{-1}}$.
Therefore, the lifetime would increase to several hundred seconds when the trapping environment is
cooled to liquid nitrogen temperatures (\unit[77]{K}). Moreover, one has to consider that many of
the final states are also confined in the ac trap and, eventually, population will also be
transferred back to the ground state. This shows that large molecules can be confined in traps once
they are decelerated to slow enough velocities and that, while thermalization due to blackbody
radiation has to be considered at room temperature, it will not prohibit trapping of large molecules
and it is negligible at liquid nitrogen temperatures.

\subsubsection{Conformer selection with the alternating-gradient decelerator}
\label{sec:AG-dec:conformer-selection}

Generally, the quantum-state selectivity of the AG decelerator provides for the possibility to
prepare clean samples of individual conformers of large molecules. Similar to the AG focuser, any
polar conformer can be addressed. Additionally, even without the deceleration to very slow
velocities the reduced velocity and the correspondingly longer flight-times to the detector allow
for a practically complete temporal separation of the accepted molecules from the remaining beam.
Therefore, the AG decelerator provides, in principle, the best selectivity of different conformers
present in an molecular beam. Decelerating a single conformer to slow velocities and successively
trapping it in ac traps provides additional selectivity and allows for the separation of quantum
states and conformers with quite similar $m/\mu$ ratios. However, this strong selectivity comes at
the price of a more complicated experimental setup.

\section{Conclusions}
\label{sec:conclusions}

We have compared and demonstrated different experimental approaches for the quantum-state selective
manipulation of the motion of polar molecules. All methods are generally applicable for molecules in
low- and high-field seeking states. However, the methods presented here are well adapted to the
manipulation of large molecules, where all quantum states are high-field seeking.

Generally, all the different techniques described allow the spatial separation of different isomers
of neutral species. While we have demonstrated this already for 3-aminophenol using the
$m/\mu$~selector, it is clear that the quantum state selectivity of the deflector and the AG
decelerator can be exploited in very much the same way. Moreover, both these alternative techniques
can provide background free samples of large molecules. Generally, each approach has its advantages
and disadvantages, which shall be summarised here.

Using the deflector one disperses quantum states of particles of identical mass according to their
effective dipole moments \mueff{}, with the most polar states deflected the most. Therefore, using
focused lasers or narrow slits for spatial discrimination of the sample, one can perform experiments
using samples of these most polar quantum states. Contrary to the AG selector, these are pure
samples without any background from unpolar states or conformers, or seed gas, as is shown by the
simulations for cis and trans 3-aminophenol in figure~\ref{fig:deflection:3AP}. In principle this
background in experiments with the AG focuser can be removed by bending or tilting the device with
respect to the molecular beam axis or by placing a beam-stop on the molecular beam axis in order to
remove particles unaffected by the electric fields, but these changes will also considerably reduce
the transmission of the selected particles. The AG focuser, however, has the advantage that it
allows the selection of any conformer whose dipole moment differs (enough) from the other
conformers.

Background reduction can, of course, also be performed in time. If one uses the alternating gradient
decelerator to slow the accepted molecular packet -- containing only a single conformer -- down to
reach the detector only after the original pulse has passed, one can also perform practically
background free experiments with the accepted packets. Of course, the ultimate experiment in that
respect would be the deceleration and subsequent trapping of a single conformer.

For future proof-of-principle experiments on single conformers the deflector is surely the most
promising setup. It is the simplest of the presented techniques, and it provides a background-free
sample of the most polar quantum states of the most polar isomer, the states that can also be
manipulated the strongest in successive laser or dc electric field experiments, like mixed field
orientation~\cite{Holmegaard:PRL102:023001}. However, if one wants to employ the separation in
routine experiments where specific or multiple isomers must be addressed individually, generally
dynamic focusing is obligatory in order to obtain enriched samples of any but the most polar
conformer. Whether the experiments requires really pure samples or not determines whether the
simpler focuser or the very complex decelerator shall be used.

Whereas the discussions of $m/\mu$ selection in this manuscript were restricted to the separation of
isomers of molecules (fixed $m$), the selection can of course also be used to separate species from
any other mixture based on the $m/\mu$-ratio, \ie, to separate polar clusters from the typically
very broad size distributions.

Generally, the clean, well-defined samples provided by these experiments could aid or even just
allow novel experiments with complex molecules, for instance, femto-second pump-probe measurements,
x-ray or electron diffraction in the gas-phase, or tomographic reconstructions of molecular
orbitals. Such samples could also be very advantageous for metrology applications, such as, for
example, matter-wave interferometry or the search for electroweak interactions in chiral molecules.

\appendix
\section{Stark energies of asymmetric rotors}
\label{sec:stark-calc}

In this section we want to summarise the details of the calculations of adiabatic energy curves for
asymmetric top molecules. The Hamiltonian matrix is set up in the basis of symmetric top
wavefunctions. For the asymmetric rotor in an electric field only $M$ is a good quantum number, as
$K$ is mixed by the molecular asymmetry and $J$ is mixed by the field. Therefore, one can treat the
different $M$ levels individually, but needs to set up the $M$ matrices including all $J$ and $K$
levels. For the accurate description of higher rotational states it is also important to include
centrifugal distortion constants. The Hamiltonian $H$ of an asymmetric rotor molecule with dipole
moment $\vec{\mu}$ in an electric field of strength $E$ might be written as the sum of the
Hamiltonian $H_\text{rot}$ of an asymmetric rotor in free space and the contribution due to the
Stark effect $H_\text{Stark}$ as
\begin{equation*}
   H = H_\text{rot} + H_\text{Stark}
\end{equation*}
Following references\hspace{1ex} \onlinecite{Watson:VibSpecStruct6:1} and\hspace{1ex}
\onlinecite{AbdElRahim:JPCA109:8507}, the corresponding matrix elements, using symmetric rotor basis
functions, representation $I^r$, and Watson's A-reduction~\cite{Watson:VibSpecStruct6:1}, are:
\begin{widetext}
   \vspace{-5mm}
   \begin{align*}
      \left<JKM \mid H_\text{rot} \mid JKM\right> &= { \frac{B+C}{2} \left(J(J+1)-K^2\right) + AK^2 } \\
      & \quad - \Delta_J J^2(J+1)^2 - \Delta_{JK} J(J+1)K^2 - \Delta_K K^4 \\
      \left<JK\pm2M \mid H_\text{rot} \mid JKM\right> &= { \left(\frac{B-C}{4} -
            \delta_J J(J+1) - \frac{\delta_K}{2}\left((K\pm2)^2 + K^2\right) \right)} \\
      & \quad \cdot \sqrt{J(J+1) - K(K \pm 1)} \sqrt{J(J+1)-(K \pm 1)(K \pm 2)} \\
      \left<JKM \mid \mu_a \mid JKM\right> &= -\frac{MK}{J(J+1)} \mu_a E \\
      \left<J+1KM \mid \mu_a \mid JKM\right> &= \left<JKM \mid \mu_a \mid J+1KM\right> \\
      &= {- \frac{\sqrt{(J+1)^2-K^2} \sqrt{(J+1)^2 - M^2}}{(J+1) \sqrt{(2J+1)(2J+3)}} \mu_a E } \\
      \left<JK\pm1M \mid \mu_b \mid JKM\right> &= {-
         \frac{M \sqrt{(J \mp K)(J \pm K+1)}}{2J(J+1)} \mu_b E }\\
      \left<J+1K\pm1M \mid \mu_b \mid JKM\right> &= \left<JK\pm1M \mid \mu_b \mid J+1KM\right>  \\
      &= {\pm \frac{\sqrt{(J \pm K+1)(J \pm K+2)}\sqrt{(J+1)^2 - M^2}}{2(J+1) \sqrt{(2J+1)(2J+3)} } \mu_b E } \\
      \left<JK\pm1M \mid \mu_c \mid JKM\right> &= {\pm i
         \frac{M \sqrt{(J \mp K)(J \pm K+1)}}{2J(J+1)} \mu_c E }\\
      \left<J+1K\pm1M \mid \mu_c \mid JKM\right> &= \left<JK\pm1M \mid \mu_c \mid J+1KM\right> \\
      &= {-i \frac{\sqrt{(J \pm K+1)(J \pm K+2)} \sqrt{(J+1)^2 - M^2 }}{2(J+1) \sqrt{(2J+1)(2J+3)}} \mu_c E }
   \end{align*}
\end{widetext}
For the correct assignment of the states to the ``adiabatic quantum number labels''
$\tilde{J}\tilde{K_a}\tilde{K_c}\tilde{M}$, \ie, to the adiabatically corresponding field-free rotor
states, one has to classify the states according to their character in the \emph{electric field
   symmetry group}~\cite{Watson:CJP53:2210, Bunker:MolecularSymmetry}. This symmetry classification
is performed by applying a Wang transformation~\cite{Wang:PR34:243} to the Hamiltonian matrix. If
the molecule's dipole moment is along one of the principal axes of inertia, the matrix will be block
diagonalised by this transformation according to the remaining symmetry in the field and the blocks
are treated independently. For arbitrary orientation of the dipole moment in the inertial frame of
the molecule the full matrix must be diagonalised. In any case, this process ensures that all states
(eigenvalues and eigenvectors) obtained from a single matrix diagonalisation do have the same
symmetry, and, therefore, no real crossings between these states can occur. Therefore, by sorting
the resulting levels by energy and assigning quantum number labels in the same order as for the
field-free states of the same symmetry yields the correct adiabatic labels.

These calculations are performed for a number of electric field strengths -- typically in steps of
1~kV/cm from 0~kV/cm to 250~kV/cm -- and the resulting energies are stored for later use in
simulations using the libcoldmol program package~\cite{Kuepper:libcoldmol}.

\bibliographystyle{fd}
\bibliography{string,mp}

\end{document}